\documentstyle{article}

\begin{document}

\title{Stable three-dimensional spinning optical solitons supported by competing
quadratic and cubic nonlinearities}
\author{}
\maketitle

\begin{center}
D. Mihalache$^1$, D. Mazilu$^1$, L.-C. Crasovan$^1$, I. Towers$^2$, B. A.
Malomed$^2$, A. V. Buryak$^3$, L. Torner$^4$, and F. Lederer$^5$\\[0pt]

\bigskip

$^1$Department of Theoretical Physics, Institute of Atomic Physics, PO Box
MG-6, Bucharest, Romania\\[0pt]

$^2$Faculty of Engineering, Tel Aviv University, Tel Aviv 69978, Israel
\\[0pt]

$^3$School of Mathematics and Statistics, University of New South Wales at
ADFA, Canberra, ACT 2600, Australia\\[0pt]

$^4$Department of Signal Theory and Communications, Universitat Politecnica
de Catalunya, ES 08034 Barcelona, Spain\\[0pt]

$^5$Institute of Solid State Theory and Theoretical Optics,
Friedrich-Schiller University Jena, Max-Wien-Platz 1, D-07743, Jena, Germany
\\[0pt]

\bigskip \bigskip

{\large {Abstract} }
\end{center}

We show that the quadratic interaction of fundamental and second harmonics
in a bulk dispersive medium, combined with self-defocusing cubic
nonlinearity, give rise to completely localized spatiotemporal solitons
(vortex tori) with vorticity $s=1$. There is no threshold necessary for the
existence of these solitons. They are found to be stable if their energy
exceeds a certain critical value, so that the stability domain occupies
about 10 {\%} of the existence region of the solitons. On the contrary to
spatial vortex solitons in the same model, the spatiotemporal ones with $s=2$
are never stable. These results might open the way for experimental
observation of spinning three-dimensional solitons in optical media.

\newpage

\section{Introduction}

Solitons, i.e., self-trapped light beams or pulses that are supported by a
balance between diffraction and/or dispersion and nonlinearity, are
prominent objects in nonlinear optics \cite{George}. Optical spatiotemporal
solitons (STS) \cite{KanRub}, alias superspikes \cite{MBA} or light bullets 
\cite{Yaron}, were predicted in many works \cite{KanRub}~-~\cite{greek}.
They result from the simultaneous balance of diffraction and group-velocity
dispersion (GVD) by self-focusing. Although they cannot be stable in the
uniform self-focusing Kerr ($\chi ^{(3)}$) medium \cite{collapse}, stability
can be achieved in saturable \cite{MBA,AS,Edmundson}, quadratically
nonlinear ($\chi ^{(2)}$) \cite{KanRub,quadr1,quadr2,quadr3}, and
graded-index Kerr media \cite{Agrawal}. STS can also be found in
off-resonance two-level systems \cite{Igor}, in self-induced-transparency
media \cite{Miriam}, as well as in engineered tandem structures
incorporating quadratically nonlinear slices \cite{tandem}.

While a fully localized ``light bullet'' in three dimensions (3D) has not
yet been found in an experiment, two-dimensional (2D) STS in a bulk $\chi
^{(2)}$ medium were observed \ in Ref. \cite{Wise}. That work reported the
formation of pulses in quadratic media, which overcome diffraction in one
transverse spatial dimension and GVD in the longitudinal direction. However,
such experiments were performed by means of the tilted-pulse technique,
which employs highly elliptical beams; therefore, diffraction is negligible
in the remaining transverse spatial dimension.

Optical vortex solitons constitute another class of self-supporting objects,
that have attracted much attention because of possible applications to the
all-optical processing of information, or to guiding and trapping of atoms.
The concepts of a multidimensional optical soliton and of an optical vortex
may be combined, giving rise to {\it spinning} (vortex) solitons. Starting
with the seminal works \cite{vortex}, both delocalized (``dark'') and
localized (``bright'') optical vortices were investigated in various 2D
environments \cite{DiTrapp,Sammut,2DvortexYuri,2DvortexSegev}. In the 3D
case, the bright spinning solitons take the shape of a torus (``doughnut'') 
\cite{Anton,new}.

For bright vortex solitons, stability is a major concern, as, unlike their
zero-spin counterparts, the spinning solitons are apt to be destabilized by
azimuthal perturbations. For 2D models with $\chi ^{(2)}$ nonlinearities, an
azimuthal instability was discovered by simulations \cite{unstable1} and
observed experimentally \cite{experiment}. As a result, a soliton with spin $
1$ splits into three or two fragments in the form of separating zero-spin
soliton. Numerical simulations of the 3D spinning STS in the $\chi ^{(2)}$
model also demonstrates splitting into moving zero-spin solitons \cite{new}.

Nevertheless, the $\chi ^{(2)}$ nonlinearity acting in combination with the
self-{\em defocusing} Kerr [$\chi _{-}^{(3)}$, where we use the subscript
``minus'' to stress the self-repulsion] nonlinearity gives rise to stable
spinning (ring-shaped) 2D solitons with spin $s=1$ and $2$ \cite{Sammut}.
Models of this type for spatial [(2+1)-dimensional] solitons are well known 
\cite{comp1,comp2}. The stability of the spinning solitons in the $\chi
^{(2)}:\chi _{-}^{(3)}$ model may be realized as a result of competition
between the self-focusing and self-defocusing nonlinearities. This
understanding is further supported by the fact that stable spinning solitons
of the same type have also been found in another optical model displaying
both focusing and defocusing nonlinearities, viz., the one based on the
cubic-quintic (CQ) nonlinear Schr\"{o}dinger (NLS) equation. In addition to
optics, the same equations have been investigated in the contexts of
Bose-Einstein condensates (BECs) \cite{Abdullaev} and Langmuir waves in
plasmas \cite{plasma} (however, in the former case, the quintic nonlinearity
arises from three-body interactions, which also give rise to losses by
recombination of BEC constituents into different species, thus making the
quintic nonlinear coefficient a complex one).

In the first direct simulations of 2D solitons with spin $1$ in the CQ
model, reported in the pioneer work \cite{Q}, it was found that they are
robust, provided that their energy is not too small \cite{Q}. Later
analysis, based on the computation of linear-stability eigenvalues,
demonstrated that some of the spinning 2D solitons considered in Ref. \cite
{Q} are subject to a weak azimuthal instability. Nonetheless, in another
part of their existence region, where they have a very large energy, the
solitons with spin $s=1$ and $s=2$ were confirmed to be stable in the 2D CQ
model \cite{IsaacPLA} (see also Ref. \cite{Berezhiani} for the stability
investigation of the solitons with spin $s=1$). Stable 2D vortex solitons in
the CQ model can self-trap from Gaussian inputs with an embedded vorticity 
\cite{Michinel}. Notice that all the solitons with $s\geq 3$ have been
demonstrated to be unstable in the CQ model \cite{IsaacPLA}.

A challenging issue is the search for physically relevant models in which 
{\it stable} 3D spinning solitons exist. In fact, the only previously known
model which could support stable 3D vortex solitons was the Skyrme model
(see reviews \cite{reviews}). Very recently, we have found stable 3D
spinning STS in the CQ model, which could again be construed as a result of
the competition between self-focusing and self-defocusing \cite{PRL}. Direct
simulations of the 3D CQ model \cite{new2} demonstrated that 3D spinning
solitons with moderate energies were unstable against azimuthal
perturbations, while the ones with very large energies, i.e., broad
``doughnuts'' with a small hole in the center, were robust under
propagation. However, a consistent stability analysis makes it necessary to
compute eigenvalues of small perturbations. By calculating the instability
growth rates, in Ref. \cite{PRL} it was rigorously shown that sufficiently
broad STS with spin $s=1$ are stable, the stability region occupying $
\approx 20\%$ of their existence region, while all the STS with $s\geq 2$
are unstable.

The aim of this paper is to show that the existence of stable spinning 3D
solitons is a more generic fact, which is not limited to the CQ model
considered in Ref. \cite{PRL}. To this end, we will analyze the existence
and stability of spinning STS solitons in the 3D version of the
above-mentioned $\chi ^{(2)}:\chi _{-}^{(3)}$ model with the self-defocusing
cubic term. In section 2, the model is formulated, and general results
concerning the existence of 3D spinning STS in it, with different values of
the spin, are displayed. Fundamental results for the stability of the
spinning solitons, based on eigenvalues found from equations linearized
around the soliton solutions, are presented in section 3. Direct simulations
of the solitons' stability within the framework of the full nonlinear
equations are displayed in section 4, and section 5 concludes the work.

\section{The model and spinning solitons}

The scaled equations describing the reversible generation of the second
harmonic (SH) from a single fundamental-frequency (FF) component $u$, in the
presence of the self-defocusing cubic nonlinearity, dispersion and
diffraction in the (3+1)-dimensional geometry, are well known \cite
{quadr1,quadr2,Sammut,comp1,comp2}: 
\begin{eqnarray}
&&i\frac{\partial u}{\partial Z}+\frac{1}{2}\left( \frac{\partial ^{2}u} {
\partial X^{2}}+\frac{\partial ^{2}u}{\partial Y^{2}}+\frac{\partial ^{2}u} {
\partial T^{2}}\right) +u^{\ast }\,v\,-(|u|^{2}+2|v|^{2})u\,=0,  \nonumber \\
&&i\frac{\partial v}{\partial Z}+\frac{1}{4}\left( \frac{\partial ^{2}v} {
\partial X^{2}}+\frac{\partial ^{2}v}{\partial Y^{2}}+\sigma \frac{\partial
^{2}v}{\partial T^{2}}\right) -\beta v+u^{2}\,-2(2|u|^{2}+|v|^{2})v\,=0.
\label{scaled}
\end{eqnarray}
Here, $T$, $X$, $Y$ and $Z$ are the normalized reduced time, transverse
spatial coordinates, and propagation distance, $u$ and $v$ are envelopes of
the FF and SH fields, and $\beta $ is a phase mismatch between the FF and SH
waves. In particular, the variables used in Eqs. (\ref{scaled}) are related
to their counterparts (to be denoted by tildes) in Ref. \cite{Sammut} as
follows: $u\equiv 2\widetilde{u}$, $v\equiv \widetilde{w}$, $\sqrt{2}\left(
x,y\right) \equiv \left( \widetilde{x},\widetilde{y}\right) $.

Equations (\ref{scaled}) assume different GVD coefficients at the two
harmonics, $\sigma $ being their ratio \cite{quadr1}, but neglects the
Poynting-vector walkoff between the harmonics, and assumes that the temporal
group-velocity mismatch between them \cite{quadr3,walking,PO} has been
compensated. On the other hand, in the case $\sigma =1$ the model possesses
an additional spatiotemporal spherical symmetry \cite{quadr1,quadr2}. Below,
we will display results for the case $\sigma =1$, assuming that the
group-velocity mismatch may be neglected in this case too.

We look for stationary solutions to Eqs. (1) in the form $u=U(r,T)\exp
\left( i\kappa Z+is\theta \right) $, $v=V(r,T)\exp \left[ 2\left( i\kappa
Z+is\theta \right) \right] $, where $\theta $ is the polar angle in the
plane $\left( x,y\right) $, $\kappa $ is a wave number shift, and the
integer $s$ is the above-mentioned spin. The amplitudes $U$ and $V$ may be
taken real, obeying the equations 
\begin{eqnarray}
&&\frac{1}{2}\left( \frac{\partial ^{2}U}{\partial r^{2}}+\frac{1}{r}\frac{
\partial U}{\partial r}-\frac{s^{2}}{r^{2}}U+\frac{\partial ^{2}U}{\partial
T^{2}}\right) -\kappa U+UV-(U^{2}+2V^{2})U=0,  \nonumber \\
&&\frac{1}{4}\left( \frac{\partial ^{2}V}{\partial r^{2}}+\frac{1}{r}\frac{
\partial V}{\partial r}-\frac{4s^{2}}{r^{2}}V+\sigma \frac{\partial ^{2}V}{
\partial T^{2}}\right) -(2\kappa +\beta )V+U^{2}-2(2U^{2}+V^{2})V=0.
\label{stat}
\end{eqnarray}

Dynamical equations (\ref{scaled}) conserve the total energy 
\begin{equation}
E~=~\int \int \int (\left| u\right| ^{2}+\left| v\right| ^{2})dXdYdT\equiv
E_{u}+E_{v},  \label{E}
\end{equation}
Hamiltonian 
\begin{eqnarray}
H &=&\frac{1}{2}\int \int \int \left\{ \left[
(|u_{X}|^{2}+|u_{Y}|^{2}+|u_{T}|^{2})+\frac{1}{4}(|v_{X}|^{2}+|v_{Y}|^{2}+
\sigma |v_{T}|^{2})\right] \right.  \nonumber \\
&&+\left. \left[ \beta |v|^{2}-(u^{\ast 2}v+u^{2}v^{\ast
})+(|u|^{4}+4|u|^{2}|v|^{2}+|v|^{4})\right] \vphantom { \left[
(|u_{X}|^{2}+|u_{Y}|^{2}+|u_{T}|^{2})+ \frac{1}{4}(|v_{X}|^{2}+|v_{Y}|^{2}+
\sigma |v_{T}|^{2})\right] }\right\} dXdYdT\,,  \label{inv2}
\end{eqnarray}
momentum (equal to zero for the solutions considered), and longitudinal
component of the orbital angular momentum $L$ \cite{AA}. The following
relations between $L$, $H$ and $E$ for a stationary spinning STS follow from
Eqs. (\ref{stat}): $L=sE $, and 
\begin{equation}
H=-\frac{1}{3}\kappa E+\frac{1}{3}\beta E_{v}-\frac{1}{3}\int \int \int
\left( |u|^{4}+4|u|^{2}|v|^{2}+|v|^{4}\right) dXdYdT.  \label{H}
\end{equation}

We have numerically found one-parameter families of stationary 3D spinning
solitons which have the shape of a doughnut with a hole (supported by a
phase dislocation) in the center. To this end, we solved numerically the
coupled system of equations (\ref{stat}) using a standard band-matrix
algorithm \cite{D-B} to deal with the corresponding two-point boundary-value
problem. We will display results for $\sigma =1$; however, we have also
found that the STS exist for all $\sigma \geq 0$, cf. Ref. \cite{quadr2},
where nonspinning STS were studied in detail for $\sigma \neq 1$.

In Fig. 1 we summarize the output of extensive numerical calculations aimed
to detect the domains of existence and stability of spinning STS. The
continuous lines border the existence domain, and the dashed line
constitutes a boundary between stable and unstable regions in the parameter
plane $(\beta ,\kappa )$. The way the stability boundary has been found will
be explained in detail in the following sections.

Shapes of three representative doughnut-like {\it stable} STS are plotted in
Fig. 2 for a fixed value of the net energy [see Eq. (\ref{E})], $E=12000$.
We see that, with the increase of the mismatch $\beta $, the energy of the
FF component of the spinning soliton increases, similar to the case of
nonspinning solitons in pure $\chi ^{(2)}$ media \cite{PO,SHT,TS,Yuri}.

Below, we present systematic results which characterize spinning STS in the
case of the zero phase matching, $\beta =0$. In Fig. 3 we plot the curves 
$\kappa =\kappa (E)$ and $H=H(E)$ for {\em both} nonspinning and spinning STS
in this case. The full and dashed lines in Fig. 3 correspond to stable and
unstable branches according to results presented below. The $s=0$ solitons
are stable according to the known Vakhitov-Kolokolov (VK) criterion, which
states that the fundamental ($s=0$) soliton branch undergoes a stability
change at the point $dE/d\kappa =0$ \cite{VK}.

A feature shared by the nonspinning and spinning solitons, as it is evident
in Fig. 3, is the absence of any finite threshold for their existence. This
is a drastic difference from the recently studied STS in the CQ model, where
well-defined thresholds were found for zero and nonzero values of the spin 
\cite{Anton,PRL}.

\section{Stability eigenvalues of the spinning solitons}

Complete understanding of the stability of solitons is provided by direct
simulations of the evolution equations (see below) together with the
analysis of Eqs. (\ref{scaled}) linearized about the stationary
spinning-soliton solution. In this section, we focus on the latter approach,
seeking for perturbation eigenmodes in a general form, 
\begin{eqnarray}
&&u(Z,r,T,\theta )-U(r,T)\exp \left[ i(s\theta +\kappa Z)\right]  \nonumber
\\
&=&f(r,T)\exp \left\{ \lambda _{n}Z+i[(s+n)\theta +\kappa Z]\right\}
+g^{\ast }(r,T)\exp \left\{ \lambda _{n}^{\ast }Z+i[(s-n)\theta +\kappa
Z]\right\} \,, \\
&&v(Z,r,T,\theta )-V(r,T)\exp \left[ 2i(s\theta +\kappa Z)\right]  \nonumber
\\
&=&p(r,T)\exp \left\{ \lambda _{n}Z+i[(2s+n)\theta +2\kappa Z]\right\}
+q^{\ast }(r,T)\exp \left\{ \lambda _{n}^{\ast }Z+i[(2s-n)\theta +2\kappa
Z]\right\} \,,
\end{eqnarray}
where $n>0$ is an arbitrary integer azimuthal index of the perturbation, 
$\lambda _{n}$ is the (complex) eigenvalue that needs to be found, and
functions $f$, $g$ and $p$, $q$ obey equations 
\begin{eqnarray}
i\lambda _{n}f+\frac{1}{2}\left[ \frac{\partial ^{2}f}{\partial T^{2}}+ 
\frac{\partial ^{2}f}{\partial r^{2}}+r^{-1}\frac{\partial f}{\partial r}
-(s+n)^{2}r^{-2}f\right] -\kappa f &&  \nonumber \\
-2\left( U^{2}+V^{2}\right) \,f-\left( U^{2}-V\right) g-\left( 2UV-U\right)
p-2UVq &=&0, \\
-i\lambda _{n}g+\frac{1}{2}\left[ \frac{\partial ^{2}g}{\partial T^{2}}+ 
\frac{\partial ^{2}g}{\partial r^{2}}+r^{-1}\frac{\partial g}{\partial r}
-(s-n)^{2}r^{-2}g\right] -\kappa g &&  \nonumber \\
-2\left( U^{2}+V^{2}\right) \,g-\left( U^{2}-V\right) f-\left( 2UV-U\right)
q-2UVp &=&0, \\
i\lambda _{n}p+\frac{1}{4}\left[ \sigma \frac{\partial ^{2}p}{\partial T^{2}}
+\frac{\partial ^{2}p}{\partial r^{2}}+r^{-1}\frac{\partial p} {\partial r}
-(2s+n)^{2}r^{-2}p\right] -(2\kappa +\beta )p &&  \nonumber \\
-4\left( U^{2}+V^{2}\right) \,p-2V^{2}q-2\left( 2UV-U\right) f-4UVg &=&0, \\
-i\lambda _{n}q+\frac{1}{4}\left[ \sigma \frac{\partial ^{2}q}{\partial
T^{2} }+\frac{\partial ^{2}q}{\partial r^{2}}+r^{-1}\frac{\partial q} {
\partial r} -(2s-n)^{2}r^{-2}q\right] -(2\kappa +\beta )q &&  \nonumber \\
-4\left( U^{2}+V^{2}\right) \,q-2V^{2}p-2\left( 2UV-U\right) g-4UVf &=&0,
\end{eqnarray}
Physical solutions must decay exponentially at $r\rightarrow \infty $. At 
$r\rightarrow 0$, $f$ and $g$ must vanish as $r^{\left| s\pm n\right| }$,
whereas $p$ and $q$ vanish as $r^{\left| 2s\pm n\right| }$.

To solve the above equations and find the eigenvalues, we used a known
numerical procedure \cite{unstable1,Akh}, which produces results presented
in Fig. 4. The most persistent unstable eigenmode is found for value of the
azimuthal index $n=2$, for both $s=1$ and $s=2$. As is seen in Fig. 4, the
instability of the soliton with $s=1$, accounted for by ${\rm Re\,} \lambda
_{2}$, disappears with the increase of $\kappa $ at a stability-change
point, $\kappa _{{\rm st}}\approx 0.04572$, and the stability region extends
up to $\kappa =\kappa _{{\rm offset}}^{{\rm \ (3D)}}\approx 0.051$,
corresponding to the upper continuous line in Fig. 1, i.e., infinitely broad
solitons (which implies that the vortex of the dark-soliton type \cite
{vortex}, that may be regarded as an infinitely broad spinning soliton, is
stable too). The relative width of the stability region is $\left( \kappa _ {
{\rm offset}}^{{\rm (3D)}}-\kappa _{{\rm st} }\right) /\kappa _{{\rm offset}
}^{{\rm (3D)}}\approx 0.1$. However, {\em no} stability region has been
found for the 3D solitons with $s=2$, as well as in the 3D model of the CQ
type, and in contrast to the 2D vortex solitons in both the $\chi
^{(2)}:\chi _{-}^{(3)}$ model with the competing quadratic and cubic
nonlinearities (the same as considered here) \cite{Sammut}, and 2D CQ model 
\cite{IsaacPLA}.

In the case when the spinning solitons are unstable, their instability is 
{\it oscillatory}; the corresponding frequency, ${\rm Im}\lambda $ [see
Figs. 4 (c) and 4(d)] is found to be, generally, on the same order of
magnitude as ${\rm Re}\lambda $ at the maximum-instability point. In the
stable region, $\kappa \geq \kappa _{{\rm st}}$, all the eigenvalues are
purely imaginary. Oscillatory instabilities of solitons, characterized by
complex eigenvalues of the corresponding non-self-adjoint linear operator,
are typical to other conservative models of nonlinear optics \cite
{osc1,osc2,osc3}.

\section{Direct simulations}

The above results were checked against direct simulations of Eqs. (\ref
{scaled}), carried out by means of the Crank-Nicholson scheme. The
corresponding system of nonlinear partial differential equations was solved
by means of the Picard iteration method \cite{Ortega}, and the resulting
linear system was handled by means of the Gauss-Seidel iterative scheme. For
good convergence we needed, typically, five Picard iterations and fifteen
Gauss-Seidel iterations. We employed a transverse grid having $121\times
121\times 91$ points, and a typical longitudinal step size was $\Delta Z=0.1$.
To avoid distortion of the instability development under the action of the
periodicity imposed by the Cartesian computational mesh, we added initial
perturbations that were mimicking random fluctuations in a real system (cf.
Ref. \cite{Edmundson}).

To illustrate the evolution of a stable 3D ``bullet'' generated by an input
in the form of a completely localized Gaussian pulse with the energy 
$E_{0}=5986$ [see Eq. (\ref{E})], into which a vortex with $s=1$ was
embedded, in Fig. 5 we show the energies of its two components vs. $Z$.
Robustness of the spinning STS is attested to by the fact that it can be
generated from a Gaussian with a nested vortex, whose shape is far from the
soliton's exact form. We see from Fig. 5 that there is a strong reshaping of
the input Gaussian, which leads to a redistribution of the energy between
the two components; some energy loss occurs, caused by emission of radiation
in the course of the formation of the stable STS. Figure 6 shows gray-scale
contour plots of the intensity and phase distribution in the FF component,
in both the input Gaussian with a nested vortex, and in the emerging
spinning STS with the vorticity $s=1$ at $Z=100$, corresponding to the same
case which was presented in Fig. 5. No further essential evolution of the
soliton was observed in this case at $Z>100$.

Typical instabilities of the spinning STS with the spin $s=1$ (in the case
when it is unstable) and $s=2$ are illustrated by Figs.~7 through 10. The
azimuthal instability breaks the unstable spinning solitons into zero-spin
ones, which fly out tangentially relative to the circular crest of the
original soliton [similar to what is known about the instability-induced
breakup of the (2+1)D spatial vortex solitons \cite{unstable1}]. Thus, the
initial internal angular momentum (spin) of the doughnut-shaped spinning
soliton is converted into the orbital momentum of the emerging nonspinning
fragments.

Analyzing a large body of numerical results, we have concluded that the
number of the emerging fragments is roughly equal to twice the original spin 
$s$. The dependence of the number of the fragments on the other parameters
is fairly weak.

It is noteworthy that, in all the cases displayed in Figs.~7 through 10 (and
in many more cases not shown here), the number of the instability-generated
fragments is exactly equal to the azimuthal index of the perturbation mode
having the largest growth rate. Thus, the full nonlinear evolution of the
unstable spinning solitons is in perfect agreement with the stability
analysis based on the linearized equations, which was presented in the
previous section.

\section{Conclusion}

In this paper, we have shown that stable bright spatiotemporal spinning
solitons (vortex tori), which were recently found in the cubic-quintic model
of a dispersive optical medium with competing self-focusing and defocusing
nonlinearities \cite{PRL}, are also possible in a model based on the
competition between the quadratic and self-defocusing cubic nonlinearities.
The solitons are stable, provided that they are broad enough (so that the
soliton's energy exceeds a certain critical value, or, in other words, the
size of the internal hole is essentially smaller than the overall size of
the soliton).

In fact, the model with the $\chi ^{(2)}:\chi _{-}^{(3)}$ (quadratic-cubic)
nonlinearity may be realized easier in real optical media than the $\chi
_{+}^{(3)}:\chi _{-}^{(5)}$ (self-focusing-cubic -- self-defocusing-quintic)
one. Possibilities for the experimental implementation of the former model
(chiefly, based on the quasi-phase-matching technique) were discussed in
Refs. \cite{Sammut,Corney-Bang,Johansen}. Note that such optical media may
be used equally well for the experimental generation of both the spatial
(2+1)-dimensional solitons (vortex cylinders) considered in Ref. \cite
{Sammut} and the 3D spatio-temporal spinning solitons (vortex tori) found in
the present work.

It is relevant to stress that the amplitude of a beam that can give rise to
a stable spinning soliton should not be specifically large: as it is evident
from Fig. 2, the necessary power is essentially the same as that which is
necessary for the existence of a nonspinning soliton. The difference from
the latter case is that the beam generating a stable spinning soliton must
be broad (its cross section and temporal width should be large), i.e., its
peculiarity is not a large power but rather large total energy.

Similar to the cubic-quintic model, only spatiotemporal solitons with spin 
$s=1$ may be stable in the present system, in contrast with the spatial
(2+1)-dimensional solitons, which may be stable in the cases $s=1$ and $s=2$,
in models of both types (on the other hand, a difference from the
cubic-quintic case is that the existence of STS in the present model is not
limited by any energy threshold). These results suggest a conclusion that
stable vortex solitons are generic objects, provided that the medium's
nonlinearity contains competing elements and the soliton's energy is large
enough; in all the known models lacking the nonlinear competition, bright
vortex solitons are subject to a strong azimuthal instability.

Lastly, one can assume that, very generally speaking, the spinning soliton
is not an absolutely stable object, but rather a metastable one. Indeed, the
energy of the spinning soliton is larger than that of its zero-spin
counterpart, hence it might be possible that a very strong initial
perturbation will provoke its rearrangement into a zero-spin soliton, the
angular moment being carried away with emitted radiation. In terms of this
consideration, it appears that the $s=1$ and $s=0$ solitons are separated by
extremely high potential barriers, which make the assumed process
practically impossible. To illustrate this point, in Fig. 11 we show the
cross sections of the $s=1$ soliton which was very strongly perturbed at the
initial point, $Z=0$ (the perturbation is $\approx 30\%$ of the soliton's
amplitude), and the result of its evolution at the point $Z=200$. For the
same case, the comparison of the distributions of the intensity and phase
inside the initial strongly perturbed soliton and the finally established
one are shown in Fig. 12 (cf. Fig. 6). As is obvious from Figs. 11 and 12,
the soliton was able to completely heal the damage, remaining a truly stable
object.

\section*{acknowledgments}

D. Mihalache, D. Mazilu and L.-C. Crasovan acknowledge support from the
Deutsche Forschungsgemeinschaft (DFG) and European Community (Access to
Research Infrastructure Action of the Improving Human Potential Program).
I.T. and B.A.M. appreciate support from the Binational (US-Israel) Science
Foundation through the grant No. 1999459, and a matching support from the
Tel Aviv University. L.T. acknowledges support by TIC2000-1010.

\newpage

\section*{Figure Captions}

Fig. 1. Domains of the existence and stability of spinning STS with spin 
$s=1$. The upper continuous curve is the existence border, corresponding to
infinitely broad (in fact, dark) solitons.

Fig. 2. Typical shapes of stable STS with $s=1$ for $E=12000$: (a) 
$\beta=-0.1$, (b) $\beta=0$, and (c) $\beta=0.2$. The labels FF and SH
pertain to the fundamental-frequency and second-harmonic components of the
soliton.

Fig. 3. The propagation constant $\kappa$ (a) and Hamiltonian $H$ (b) of the
three-dimensional solitons, with different values of spin, vs. their energy 
$E$, in the case of zero phase mismatch, $\beta=0$.

Fig. 4. The growth rate of perturbations, ${\rm Re}~ \lambda$, corresponding
to different values of the azimuthal index $n$ (indicated by labels near the
curves) vs. the soliton's wave number $\kappa$: (a) $s=1$; (b) $s=2$. The
imaginary part of the stability eigenvalue, ${\rm Im}~ \lambda$,
corresponding to different values of the azimuthal index $n$ (indicated by
labels near the curves) vs. the soliton's wave number $\kappa$: (c) $s=1$;
(d) $s=2$. Here and in the following plots, $\beta=0$. We stress that, in
the case $s=1$, the instability growth rate vanishes at the point 
$\kappa=\kappa_{{\rm st}}$, see the text, while in the case $s=2$ the growth
rate corresponding to $n=2$ remains positive up to the border of the
existence region of the solitons. This border is marked in all the panels by
vertical arrows.

Fig. 5. Evolution of the energy components $E_u$ and $E_v $ of the soliton
with $s=1$, as generated by an input configuration in the form of a Gaussian
with a nested vortex. Here, the input total energy is $E=5986$.

Fig. 6. The formation of the soliton with spin $s=1$ in the same case which
corresponds to Fig. 5, shown in terms of the cross section of the fields at 
$T=0$: (a) the intensity distribution in the initial Gaussian with a nested
vortex; (b) its phase field; (c) the intensity distribution of the spinning
soliton at $Z=100$; (d) the phase field at $Z=100$.

Fig. 7. Isosurface plots illustrating the fragmentation of the $s=1$ soliton
with $\kappa=0.01$ into zero-spin ones as a result of the azimuthal
instability: (a) $Z=0$; (b) $Z=1000$.

Fig. 8. The same as in Fig. 7 in the case $\kappa =0.032$: (a) $Z=0$; (b) 
$Z=1140$.

Fig. 9. The same as in Figs. 7 and 8 in the case of the $s=2$ initial
soliton with $\kappa=0.015$: (a) $Z=0$; (b) $Z=900$.

Fig. 10. The same as in Fig. 9 in the case $\kappa =0.04$: (a) $Z=0$; (b) 
$Z=2100$.

Fig. 11. Cross sections of an $s=1$ soliton that was strongly perturbed at 
$Z=0$, and the result of its evolution after having passed the propagation
distance $Z=200$.

Fig. 12. The recovery of the soliton with spin $s=1$ in the same case as in
Fig. 11, shown in terms of the cross section of the fields at $T=0$ (cf.
Fig. 6): (a) the intensity distribution in the initial strongly perturbed
soliton; (b) its phase field; (c) the intensity distribution of the
self-cleared soliton at $Z=100$; (d) the phase field at $Z=100$.


\newpage

\begin{thebibliography}{99}
\bibitem{George}  G. I. Stegeman, D. N. Christodoulides, and M. Segev, IEEE
J. Select. Top. Quant. Electron. {\bf 6}, 1419 (2000).

\bibitem{KanRub}  A. A. Kanashov and A. M. Rubenchik, Physica D {\bf 4}, 122
(1981).

\bibitem{MBA}  J. T. Manassah, P. L. Baldeck, and R. R. Alfano, Opt. Lett. 
{\bf 13}, 1090 (1988); J. T. Manassah, {\it ibid.} {\bf 16}, 563 (1991).

\bibitem{Yaron}  Y. Silberberg, Opt. Lett. {\bf 15}, 1282 (1990).

\bibitem{Blagoeva}  A. B. Blagoeva {\it et al.}, IEEE J. Quant. Electr. {\bf 
QE-27}, {\bf \ }2060 (1991).

\bibitem{AS}  N. Akhmediev and J. M. Soto-Crespo, Phys. Rev. A {\bf 47},
1358 (1993).

\bibitem{DarkBullet}  Y. Chen and J. Atai, Opt. Lett. {\bf 20}, 133 (1995).

\bibitem{collapse}  L. Berg{\'{e}}, Phys. Rep. {\bf 303}, 260 (1998).

\bibitem{Enns}  R. H. Enns {\it et al.}, Opt. Quant. Electron. {\bf 24},
S1295 (1992); R. McLeod, K. Wagner, and S. Blair, Phys. Rev. A {\bf 52},
3254 (1995); J. T. Manassah and B. Gross, Laser Phys. {\bf 7}, 9 (1997).

\bibitem{Edmundson}  D. E. Edmundson, Phys. Rev. E {\bf 55}, 7636 (1997).

\bibitem{Skarka}  V. Skarka, V. I. Berezhiani, and R. Miklaszewski, Phys.
Rev. E {\bf 56}, 1080 (1997).

\bibitem{quadr1}  K. Hayata and M. Koshiba, Phys. Rev. Lett. {\bf 71}, 3275
(1993); B. A. Malomed {\it et al.}, Phys. Rev. E {\bf 56}, 4725 (1997); D.
V. Skryabin and W. J. Firth, Opt. Commun. {\bf 148}, 79 (1998); D.
Mihalache, D. Mazilu, B. A. Malomed, and L. Torner, {\it ibid.} {\bf 152},
365 (1998).

\bibitem{quadr2}  D. Mihalache, D. Mazilu, J. D\"{o}rring, and L. Torner,
Opt. Commun. {\bf 159}, 129 (1999).

\bibitem{quadr3}  D. Mihalache, D. Mazilu, B. A. Malomed, and L. Torner,
Opt. Commun. {\bf 169}, 341 (1999); D. Mihalache, D. Mazilu, L.-C. Crasovan,
L. Torner, B. A. Malomed, and F. Lederer, Phys. Rev. E {\bf 62}, 7340 (2000).

\bibitem{Agrawal}  S. Raghavan and G. P. Agrawal, Opt. Commun. {\bf 180},
377 (2000).

\bibitem{greek}  H. E. Nistazakis, D. J. Frantzeskakis, and B. A. Malomed,
Phys. Rev. E {\bf 64}, 026604 (2001); C. Polymilis, D. J. Frantzeskakis, A.
N. Yannacopoulos, K. Hizanidis, and G. Rowlands, J. OPt. Soc. Am. B {\bf 18}
, 75 (2001).

\bibitem{Igor}  I. V. Mel'nikov, D. Mihalache, and N.-C. Panoiu, Opt.
Commun. {\bf 181}, 345 (2000).

\bibitem{Miriam}  M. Blaauboer, B. A. Malomed, and G. Kurizki, Phys. Rev.
Lett. {\bf 84}, 1906 (2000).

\bibitem{tandem}  L. Torner, S. Carrasco, J. P. Torres, L.-C. Crasovan, and
D. Mihalache, Opt. Commun. {\bf 199}, 277 (2001).

\bibitem{Wise}  X. Liu, L. J. Qian, and F. W. Wise, Phys. Rev. Lett. {\bf 82},
4631 (1999); X. Liu, K. Beckwitt, and F. Wise, Phys. Rev. E {\bf 62}, 1328
(2000).

\bibitem{vortex}  G. A. Swartzlander, Jr., and C. T. Law, Phys. Rev. Lett. 
{\bf 69}, 2503 (1992); A. W. Snyder, L. Poladian, and D. J. Mitchell, Opt.
Lett. {\bf 17}, 789 (1992).

\bibitem{DiTrapp}  P. Di Trapani {\it et al.}, Phys. Rev. Lett. {\bf 84},
3843 (2000).

\bibitem{Sammut}  I. Towers, A.V. Buryak, R.A. Sammut, and B.A. Malomed,
Phys. Rev. E {\bf 63}, 055601(R) (2001).

\bibitem{2DvortexYuri}  A. S. Desyatnikov and Yu. S. Kivshar, Phys. Rev.
Lett. {\bf 87}, 033901 (2001).

\bibitem{2DvortexSegev}  Z. H. Musslimani, M. Soljacic, M. Segev, and D. N.
Christodoulides, Phys. Rev. E {\bf 63}, 066608 (2001).

\bibitem{Anton}  A. Desyatnikov, A. Maimistov, and B. Malomed, Phys. Rev. E 
{\bf 61}, 3107 (2000).

\bibitem{new}  D. Mihalache, D. Mazilu, L.-C. Crasovan, B. A. Malomed, and
F. Lederer, Phys. Rev. E {\bf 62}, R1505 (2000).

\bibitem{unstable1}  L. Torner and D. V. Petrov, Electr. Lett. {\bf 33}, 608
(1997); W. J. Firth and D. V. Skryabin, Phys. Rev. Lett. {\bf 79}, 2450
(1997); J. P. Torres, J. M. Soto-Crespo, L. Torner, and D. V. Petrov, J.
Opt. Soc. Am. B {\bf 15}, 625 (1998); D. V. Skryabin and W. J. Firth, Phys.
Rev. E {\bf 58}, 3916 (1998).

\bibitem{experiment}  D. V. Petrov {\it et al.}, Opt. Lett. {\bf 23}, 1444
(1998).

\bibitem{comp1}  A. V. Buryak, Yu. S. Kivshar, and S. Trillo, Opt. Lett. 
{\bf 20}, 1961 (1995), M. A. Karpierz, Opt. Lett. {\bf 20}, 1677 (1995).

\bibitem{comp2}  O. Bang, J. Opt. Soc. Am. B {\bf 14}, 51 (1997); O. Bang,
Yu. S. Kivshar, and A. V. Buryak, Opt. Lett. {\bf 22}, 1680 (1997); L.
Berge, O. Bang, J. Juul Rasmussen, and V. K. Mezentsev, Phys. Rev. E {\bf 55},
555 (1997); O. Bang, Yu. S. Kivshar, A. V. Buryak, A. De Rossi, and S.
Trillo, Phys. Rev. E {\bf 58}, 5057 (1998).

\bibitem{Abdullaev}  F. Kh. Abdullaev, A. Gammal, L. Tomio, and T.
Frederico, Phys. Rev. A {\bf 63}, 043604 (2001); A. Gammal, T. Frederico, L.
Tomio, and F. Kh. Abdullaev, Phys. Lett. A {\bf 267}, 305 (2001).

\bibitem{plasma}  C. T. Zhou and X. T. He, Physica Scripta, {\bf 50}, 415
(1994).

\bibitem{Q}  M. Quiroga-Teixeiro and H. Michinel, J. Opt. Soc. Am. B {\bf 14},
2004 (1997).

\bibitem{IsaacPLA}  I. Towers {\it et al.}, Phys. Lett. A {\bf 288}, 292
(2001); B. A. Malomed, L.-C. Crasovan, D. Mihalache, Physica D {\bf 161},
187 (2002); R. L. Pego and H. A. Warchall, Los Alamos e-print archive:
nlin.PS/0108009.

\bibitem{Berezhiani}  V. I. Berezhiani, V. Skarka, and N. B. Aleksic, Phys.
Rev. E {\bf 64}, 057601 (2001).

\bibitem{Michinel}  H. Michinel {\it et al.} J. Opt. B: Quantum Semiclass.
Opt. {\bf 3}, 314 (2001).

\bibitem{reviews}  D. O. Riska, Adv. Nucl. Phys. {\bf 22}, 1 (1996); T.
Gisiger and M. B. Paranjape, Phys. Rep. {\bf 306}, 110 (1998).

\bibitem{PRL}  D. Mihalache, D. Mazilu, L.-C. Crasovan, I. Towers, A. V.
Buryak, B. A. Malomed, L. Torner, J. P. Torres, and F. Lederer, Phys. Rev.
Lett. {\bf 88}, 073902 (2002).

\bibitem{new2}  D. Mihalache, D. Mazilu, L.-C. Crasovan, B. A. Malomed, and
F. Lederer, Phys. Rev. E {\bf 61}, 7142 (2000).

\bibitem{walking}  L. Torner, D. Mazilu, and D. Mihalache, Phys. Rev. Lett. 
{\bf 77}, 2455 (1996); C. Etrich, U. Peschel, F. Lederer, and B. A. Malomed,
Phys. Rev. E {\bf 55}, 6155 (1997).

\bibitem{PO}  C. Etrich, F. Lederer, B. A. Malomed, T. Peschel, and U.
Peschel, Progress in Optics {\bf 41}, 483 (2000).

\bibitem{AA}  N. N. Akhmediev and A. Ankiewicz, {\it Solitons, Nonlinear
Pulses and Beams} (Chapman and Hall, London, 1997).

\bibitem{D-B}  {\ G. Dahlquist and {\AA }. Bj{\"{o}}rk, {\it Numerical
Methods}, (Prentice Hall, Englewood Cliffs, 1974). }

\bibitem{SHT}  G. I. Stegeman, D. J. Hagan, and L. Torner, Opt. Quantum
Electron. {\bf 28}, 1691 (1996).

\bibitem{TS}  L. Torner and A. P. Sukhorukov, Opt.\ Photonics News {\bf 13},
26 (2002).

\bibitem{Yuri}  Yu. S. Kivshar, in {\it Advanced Photonics with Second-Order
Optically Nonlinear Processes}, edited by A. D. Boardman, L. Pavlov, and S.
Panev (Kluwer Academic, Dordrecht,1998), p. 451.

\bibitem{VK}  M. G. Vakhitov and A.A. Kolokolov, Radiophys. Quantum El. {\bf
16}, 783 (1973).

\bibitem{Akh}  J. M. Soto-Crespo, D. R. Heatley, E. M. Wright, and N. N.
Akhmediev, Phys. Rev. A {\bf 44}, 636 (1991); J. Atai, Y. Chen, and J. M.
Soto-Crespo, Phys. Rev. A {\bf 49}, R3170 (1994).

\bibitem{osc1}  N. N. Akhmediev, A. Ankiewicz, and H. T. Tran, J. Opt. Soc.
Am. B {\bf 10}, 230 (1993).

\bibitem{osc2}  B. A. Malomed and R. S. Tasgal, Phys. Rev. E {\bf 49}, 5787
(1994); I. V. Barashenkov, D. E. Pelinovsky, and E. V. Zemlyanaya, Phys.
Rev. Lett. {\bf 80}, 5117 (1998); A. De Rossi, C. Conti, and S. Trillo,
Phys. Rev. Lett. {\bf 81}, 85 (1998); J. Sch\"{o}llmann, R. Scheibenzuber,
A. S. Kovalev, A. P. Mayer, and A. A. Maradudin, Phys. Rev. E {\bf 59}, 4618
(1999).

\bibitem{osc3}  D. Mihalache, D. Mazilu, and L. Torner, Phys. Rev. Lett. 
{\bf 81}, 4353 (1998); D. Mihalache, D. Mazilu, and L.-C. Crasovan, Phys.
Rev. E {\bf 60}, 7504 (1999).

\bibitem{Ortega}  J. M. Ortega, W. C. Rheinboldt, {\it Iterative Solution of
Nonlinear Equations in Several Variables}, (Academic Press, New York, 1970),
p. 182.

\bibitem{Corney-Bang}  J. F. Corney and O. Bang, Phys. Rev. E {\bf 64},
047601 (2001).

\bibitem{Johansen}  S. K. Johansen, S. Carrasco, L. Torner, O. Bang, Opt.
Commun. {\bf 203}, 393 (2002).
\end{thebibliography}
\end{document}